# QoS-aware Link Scheduling Strategy for Data Transmission in SDVN

Yong Zhang [a,b,c], Mao Ye [c,*], Lin Guan [c]

[a] Institute of Advanced Technology on Communication, Chengdu Technological University, Chengdu, China

[b] National Key Laboratory of Science and Technology on Communication, University of Electronic Science and Technology of China, Chengdu, China

[c] Department of Computer Science, Loughborough University, Loughborough, UK

**Abstract**

The vehicular ad-hoc network (VANET) based on dedicated short-range communication (DSRC) is a distributed communication system, in which all the nodes share the wireless channel with carrier sense multiple access/collision avoid (CSMA/CA) protocol. However, the backoff mechanism of CSMA/CA in the channel contention might cause uncertain transmission delay and impede a certain quality of service (QoS) of applications. Moreover, there still exists a possibility of parlous data-packets collision, especially for broadcast or non-acknowledgement (NACK) transmissions. The original contributions of this paper are summarized as follows: (1) Model the packets collision probability of broadcast or NACK transmission in VANET with the combination theory and investigate the potential influence of miss my packets (MMP) problem. (2) Based on the software define vehicular network (SDVN) framework and QoS requirement, a novel link-level scheduling strategy, which determines the start-sending time for each connection, is proposed to maximize packets delivery ratio (PDR). Alternatively, maximizing PDR has been converted to the overlap minimization among transmission durations. (3) Meanwhile, an innovative transmission scheduling greedy search (TSGS) algorithm is originally proposed to mitigate computational complexity. Extensive simulations have been done in a unified platform Veins combining SUMO and OMNET++. And numerous results show that the proposed algorithm can effectively improve the PDR by at least 15%, enhance the collision-avoidance performance by almost 40%, and reduce the MMP ratio by about 3% compared with the random transmitting, meanwhile meet the QoS requirement.

## 1. Introduction

Vehicular ad-hoc network (VANET) with inter-vehicular communication has been an emerging technology for efficient safety and entertainment information dissemination. However, highly dynamic VANET topology due to the rapid mobility and varying vehicle density can cause unstable wireless links, and further lead to more packets loss and increased transmission delay, which bring challenges to guarantee the quality of service (QoS). The dedicated short-range communication (DSRC) [1] is a special wireless technology aiming at supporting both vehicle-to-infrastructure (V2I) and vehicle-to-vehicle (V2V) communications. DSRC also refers to as a suite of standards including IEEE 802.11p, IEEE 1609.1/.2/.3/.4 protocol family and SAE J2735 message set dictionary[2]. DSRC is essentially a distributed communication system, and all connections in the system share the wireless channel, that is why a competition mechanism, carrier sense multiple access with collision avoidance (CSMA/CA), is adopted here. Although the distribution nature of DSRC can adapt to the rapidly changing topology in VANET, DSRC still faces many challenges due to a lack of global information, such as efficient resource utilization, effective information dissemination, precise time-delay control, etc. Software defined network (SDN) with the centralized characteristic paves a beneficial way for network configuration, resource management and performance improvement in VANET. Therefore, the convergence of SDN with VANET, called software define vehicular network (SDVN), has gained significant attention from both academia and industry[3],[4],[5],[6],[7],[8]. Because the QoS requirements are important for VANET safety, emergency, and multimedia services, there are numerous studies investigating QoS in SDVN[9], [10], [11], [12], [13] which mainly involve the upper layer protocol.

* Corresponding author.

Besides, the improvement based on the lower layer protocol like the media access control (MAC) layer, is another research direction to enhance the VANET performance. And compared with that of the upper layer protocol, the lower-layer related improvement has more advantages of high efficiency. In the DSRC system, CSMA/CA is a basic protocol of MAC layer for the sharing wireless channel within nodes. There have been many studies focused on the performance improvement of CSMA/CA by adjusting its contention window (CW) [14],[15],[16]. In addition to directly improving CSMA/CA protocol, the combination of time division multiple access (TDMA) and MAC layer to form the overlay protocol is another optimization direction of MAC layer in VANET[17], [18], [19], [20], [21], [22], [23].

In this paper, under the SDVN framework, we propose a new approach for the link-level scheduling to improve delivery performance of VANET meanwhile meet QoS requirement in term of the transmission delay, in which a link consists of a sending node and a receiving node. Each transmission is also known as a connection which can send one or multiple data packets. The connection here is the scheduling unit of the proposed algorithm, therefore, our research is based on the lower layer protocols. However, different from the previous TDMA-based overlay, our approach avoids the allocation and synchronization issues of time slots. The contributions of this work are outlined as stated below:

- Model the packets-collision probability of broadcast or non-acknowledgement (NACK) transmission in VANET with combinatorial theory (combinations with/without repetition) and consider the probable existence of miss my packets (MMP) problem.

- Based on SDVN framework, a novel scheduling strategy is proposed to schedule the start-sending time of each transmission to maximize the packet delivery ratio (PDR) meanwhile meet QoS about the transmission delay, and avoid the issues of time-slot allocation and synchronization.

- An innovative transmission scheduling greedy search (TSGS) algorithm is proposed to avoid high computational complexity.

The remainder of this paper is organized as follows: Section 2 briefly describes the related work for QoS in SDVN and MAC optimization in VANET. The problem description is presented in Section 3. System framework and proposed algorithm are explained in Section 4. The numerical experiments are displayed in Section 5. In Section 6, we draw concluding remarks.

## 2. Related work

This section presents some researches related to our work. Our work is about QoS-driven MAC layer optimization built on the VANET and SDN framework. Therefore, relative work on both SDVN and QoS will be introduced first to declare the pluses of SDN framework when handling QoS issues in VANET. However, there still exist many minuses in existing researches for MAC layer optimization. Some existing work on MAC layer optimization will then be summarized to show the background and some QoS-related limitations. And our work aims to bridge the gap and accentuate the optimization on the lower layer protocols of VANET by involving the SDN framework.

*2.1 SDVN and QoS*

SDN, first proposed in [3], is a paradigm based on the decoupling of the data forwarding plane and the control plane. Nowadays, SDN has been widely introduced into many fields such as internet of things ( IoT)[24], VANET[4], etc. [4] introduced SDN to VANET field by presenting a data dissemination system via the hybrid of V2I and V2V communications and discussing the scheduling challenges arising in such an environment. An online scheduling algorithm was proposed to enhance the data-dissemination performance by exploiting the synergy between V2I and V2V communications. Authors introduced the SDN concept and enabled the centralized scheduling at the roadside unit (RSU), which resembles the concept of software defined network in VANET. With the introduction of SDN controllers in software define vehicular network (SDVN), new ideas can be introduced into the VANET for solving important traditional issues such as data packet delivery routing [5],[6],[7], less congested path searching[8], etc.

Besides, SDVN can be considered as an information transmission network in some ways, where QoS is still an important concern, and the issue of QoS is considered in numerous SDVN studies. [9] proposed a traffic differentiated clustering routing (TDCR) mechanism under SDN architecture, in which an SDN controller takes charge of RSUs and a base station (BS). The proposed solution involves the access (DSRC or cellular vehicular network) selection to reduce the cellular bandwidth cost while guaranteeing the QoS in the aspect of end to end delay. [10] introduced a link-connectivity aware routing framework for a general SDVN. The proposed routing protocol makes use of unicast, broadcast, and store-carry-forward concepts. The resulting tribrid routing framework targets on finding the shortest and stable enough route to deliver a given set of packets with the expected QoS in terms of latency. To reliably support infotainment and other video streaming related applications with a certain QoS, [11] described a possible approach to tackle this issue based on combining three paradigms: content-centric networking (CCN), floating content (FC) and SDN. [12] proposed a combination scheme to support network virtualization in software defined VANET. Two solutions, encryption isolation and overlay isolation, are chosen to isolate virtual networks. Authors combine these two solutions and optimize the solution assignment to guarantee the QoS of each virtual VANET. Considering the QoS of vehicular applications, [13] proposed a combination of an SDN approach and a fog computing architecture for future 5G-VANET systems. By regarding vehicles as fog infrastructures and integrating them with local SDN controllers, the system is expected to become more efficient in the aspect of computation time and communication delays.

*2.2 MAC layer optimization in VANET*

As mentioned above, most of recent SDVN optimization studies for meeting QoS requirements are based on advancing the system architecture and upper-layer protocols. Meanwhile, because of the adoption of CSMA/CA in IEEE 802.11p standard, there are high possibilities of packet collisions and unbounded delay in VANET. Therefore, for guaranteeing QoS-required packet delivery, there are also sufficient studies in the scope of media access control (MAC) layer protocol to improve the performance of VANET.

Herein, it is one study direction to combine the time-division multiple access (TDMA) and CSMA/CA for the configuration of MAC in VANET. [17] proposed a demand-adaptive MAC (DAM) protocol based on dynamic TDMA and CSMA/CA. Once vehicles enter into the RSU's management domain (DM), they should shift to the listen status until getting the control message from the master. After broadcasting control messages, the vehicle could apply for a limited amount of time slots according to its demand. After the vehicle gets the authorization, it would send the message in its following time slots. Time is divided into periods called "round", and the duration time of each round could be diverse from each other. Each round has four stages: tell stage, ctrl stage, apply stage and information stage. On the basis of changeable network traffic, RSU can expand or shrink the duration of time slots dynamically round by round.

[18] presented an application-oriented TDMA overlay MAC protocol by introducing an existing RA-TDMA[19] into the vehicular context to manage state sharing in the scope of collaborative applications. Each group of vehicles engaged in one collaborative application, e.g., a platoon or smart intersection, can form a team. RA-TDMA is then used to manage the periodic transmissions of each vehicle in the team for state sharing, and allocate them to different slots in a TDMA round with a fixed application-dependent period.

[20] considered that the existing approaches with bounded access delays cannot ensure a reliable broadcast service for safety applications and proposed an approach called QoS-aware centralized hybrid MAC (QCH-MAC). The proposal tries to combine TDMA and enhanced distributed channel access (EDCA) mechanisms together into a single architecture to enhance their capabilities and to provide a higher QoS level. QCH-MAC is based on the idea that the access time can be divided into two periods: a transmission period (TP) and a reservation period (RP). TP consists of a set of TDMA slots called Tslot, while the TP period uses the EDCA protocol with two classes. The RP period is only employed by new vehicles to reserve their time slots in Tslot set. Different from IEEE 802.11p standard, the approach only considered two traffic classes: class1 and class2.

[21] proposed and verified empirical models of the relationship between network metrics and wireless conditions. Sufficient simulations were conducted with varying platoon sizes, number of occupied lanes and transmit power to deduce the empirical models. Three different MAC protocols were involved in simulations, including the native CSMA/CA MAC of IEEE 802.11p standard and two overlay TDMA protocols (referred to as PLEXE-slotted and RA-TDMAp[23]) that can be readily implemented on IEEE 802.11p standard.

In [22], the proposed communication protocol for platooning is also based on the IEEE 802.11p standard and TDMA. The rationale behind this protocol is to reduce random channel contention by adding synchronization among nodes. To decide when to send messages for nodes, authors exploited the vehicle's position within the platoon. The idea is to divide the time after a beacon from the leader into slots, and have each vehicle send its beacon in the time slot corresponding to its position in the platoon. This is different from a standard TDMA approach, as with TDMA every node participating in the communication obeys the same rules, in here, only nodes within a platoon cooperate in a TDMA-fashion.

In [23], the authors focused on the specific case of vehicles platooning applications and investigated the use of the RA-TDMA framework[19] on the top of IEEE 802.11p standard to combine the benefits of both TDMA and CSMA/CA paradigms, namely collisions reduction through synchronization of beacons and efficient bandwidth usage with asynchronous access. The authors took inspiration from the technique used in [22] in which the leader of each platoon, only, transmits its beacons at high power while the other platoon members use lower power beacons and forward information of each other in a multi-hop scheme. Similarly, [25] described a reliable self-organising TDMA (S-TDMA) and analyzed its performance with the cellular-V2X scheduler. Different from IEEE802.11p standard, this work applied S-TDMA to the long-term-extension (LTE) V2X channel structure, and it well supported V2X safety-critical transmission over the sidelink.

Except combining the TDMA and CSMA/CA to constitute overlay protocols, another study direction is to straight improve CSMA/CA mechanism. In IEEE 802.11p standard, the access category (AC) queues are applied to dispatch the priority according to various applications. Packets wait for an arbitrary inter-frame space (AIFS) before proceeding to the backoff stage. At the backoff stage, packets wait a random number of timeslots depending on the contention window (CW) size of the backoff phase and the availability of the medium, before it is transmitted to the medium. Considering the CSMA/CA employs the conventional binary exponential backoff (BEB) scheme without consideration of medium's status, [14] proposed a new adaptive backoff algorithm for the EDCA operation in IEEE 802.11p standard. The proposed algorithm takes into account the current status of the communication medium, evaluates the congestion level in the medium and uses this information to estimate the size of the backoff stage in the next transmission attempt. Similar to [14], [15] proposed an adaptive backoff algorithm based on the number of vehicle nodes, which can dynamically adjust the size of backoff window for improving the success rate of frame transmission in a unit time slot. [16] also presented a reverse back-off mechanism to adjust the size of CW instead of beginning with a small CWmin that would be increased after every failed transmission as in IEEE 802.11p standard, the reverse back-off mechanism started with a fairly large initial contention window and decreased it for every expired beacon. According to the IEEE 802.11p standard, AIFS values are fixed and deterministic. However, [26] considered that the fixed values of AIFSs do not guarantee that the higher priority AC will transmit over the medium before the lower priority AC, and proposed two algorithms to select AIFS values. The first algorithm guarantees the strict priority to higher-priority ACs despite the value of CW. It eliminates the probability of one case that a lower priority AC preempts a higher priority AC due to fluctuated AIFS values. The second algorithm is an adaptive non-deterministic algorithm that adjusts AIFS values in accordance with the current status of the medium.

As described above, there have been many studies focused on the QoS issue in SDNV, which also developed overlay protocols based on TDMA and CSMA/CA to improve performance of VANET. Also, some studies combine SDN and lower layer protocols together. For example, [27] proposed a hierarchical architecture to manage physical resources based on SDN, namely sdnMAC, in which the controller can decide the sharing range of time slot information for each RSU.

In contrast to prior work, this research, under SDVN framework, pay more attention to the improvement of lower layer protocol by overlaying MAC layer, meanwhile the allocation and synchronization of time slots are unnecessary to be involved in the new approach.

## 3. Problem Description

In this section, we present the theoretical analysis of collision issue in CSMA/CA which has worked as the channel-sharing mechanism in DSRC. The workflow of CSMA/CA in the broadcast mode is described here and the probability of data packet collision is derived with combination theory in the different number of nodes situation. Meanwhile, miss my packets (MMP) is also introduced to describe a potential problem when the receiver has only one physical receiving unit.

**3.1 Collision in CSMA/CA**

In VANET, DSRC can be adopted to establish connections among vehicles in which physical layer and media access control (MAC) layer have been defined in IEEE 802.11p standard. Since the DSRC is a distributed communication system, all connections share wireless channel by contending, and CSMA/CA protocol is introduced below.

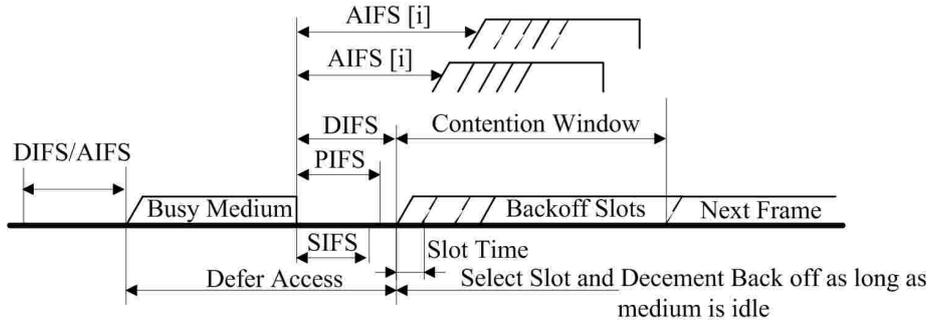

**Fig.1 The demonstration** of CSMA/CA(Obtained from [28])

As shown in Fig.1, during the packet transmission process via the medium, there must be an interval named arbitration interframe space (AIFS), which is defined as[28]

$$AIFS[AC] = AIFSN[AC] \times aSlotTime + aSIFSTime \quad (1)$$

where AC means access category with value range 0~3. Four ACs as shown in Tab.1, which can provide differentiated treatments over the wireless network, are assigned to various QoS-aware vehicles to contend for the channel access. The formula (1) presents that each AC has one distinct AIFS.

Tab.1 Access Category Coding and Description

| AC | Description |
|---|---|
| 0 | Best effort |
| 1 | Background |
| 2 | Video |
| 3 | Voice |

AIFS-number (AIFSN) indicates the number of slots which a vehicle should defer before either invoking a backoff or starting a transmission after a short interframe space (SIFS). The minimum value of the AIFSN is 2. aSlotTime defines the duration of a slot, and aSIFSTime indicates the length of SIFS. Also, the AC with a smaller AIFSN will have a higher priority to access the channel.

After the AIFS, a backoff stage will be invoked. The backoff procedure is characterized by the contention window (CW) size which is an amount of time divided into multiple time slots (TS). And CW ranges from CWmin=15 TS to CWmax=1023 (as per the standard [28]). The backoff procedure is implemented by using the Binary Exponential Backoff (BEB) strategy which has been explained in [14], however, we have made some revisions to the description as follows:

- A packet ready for transmission should wait for a certain number of TS in the backoff state.
- The original size of the backoff period is randomly selected in the 0 TS to $2^{BE}$-1 TS range, where BE is the Backoff Exponential and $2^{BE}$ can be denoted as $w$.
- In the first backoff stage, the CW is equal to CWmin. A random value, $k$, is chosen in the range of [0, CWmin], which presents the actual size of the backoff stage.
- The backoff stage is associated with a backoff counter whose initial value is equal to $k$. During each TS period, the vehicle will sense the medium status. The backoff counter will be paused if the medium is busy, otherwise it will be decremented by one.
- Once the backoff counter reaches zero, the packet will be transmitted immediately.

If the sender detects that a collision occurs, for example, the sender does not receive an expected acknowledgement (ACK) message from the receiver, the CW will be actually doubled and the size of backoff stage will be chosen in the range of [0, $2^{BE+1}$-1]. When multiple senders send packets at the same time via the medium, the collisions will occur, and then the size of backoff stage will increase.

In this paper, we analyze the broadcast collision of VANET in CSMA/CA procedure. The broadcast mode can be widely used in safety-critical information distribution, advertising, etc. It can also be applied to point-to-point information transmission without feedback. The receiving nodes will not send acknowledgements (ACKs) in the broadcast situation, so a sender never knows if anyone has received the transmitted packet correctly or not. Therefore, it's quite important to analyze CSMA/CA performance, especially the collision in this situation.

Meanwhile, it's worth mentioning that in a broadcast situation, the receiving nodes will not send ACKs. Thus, the sender performs at most one backoff, which will occur when the initial channel-access attempt senses a busy channel. Due to this, broadcast packets will never experience multiple backoffs, and the contention window will always be CWmin[29].

We assume that there are $n$ nodes (or vehicles) in VANET, the CSMA/CA broadcast Markov model can be demonstrated as Fig.2.

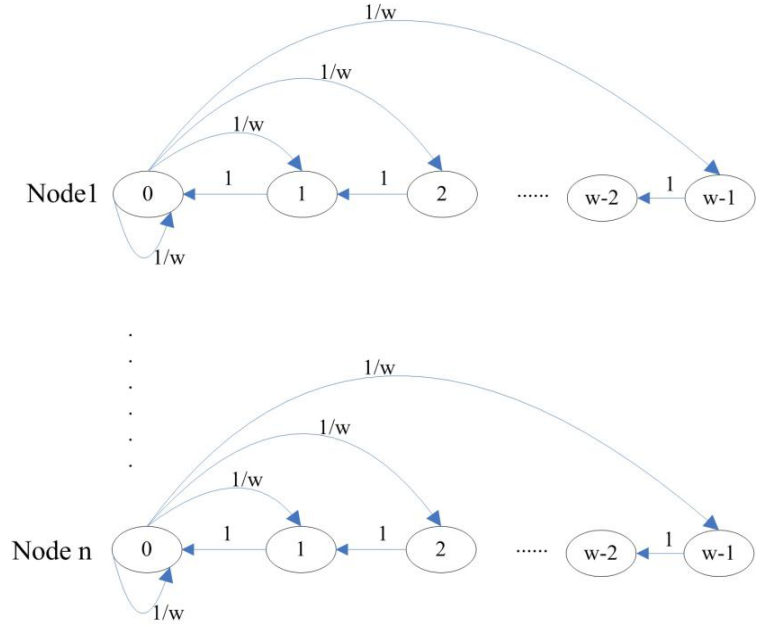

**Fig.2 CSMA/CA broadcast Markov model**

When vehicles generate new packets for broadcasting, the backoff time of each packet is randomly selected from a contention-window range of 0 to $w-1$ slot times. As mentioned before, the backoff time (counter) will be decremented every slot time when the channel is sensed idle. When the counter reaches 0, the vehicle will transmit the packet immediately. If the channel is determined to be busy, the counter will be paused. Assuming that vehicles conduct the backoff process correctly (e.g., no hidden node problem) in Fig.2, the state transition probabilities can be given by[30]

$$\begin{cases} P\{k \mid 0\} = \dfrac{1}{w}, & \text{for } k \in [0, w-1] \\ P\{k-1 \mid k\} = 1, & \text{for } k \in [1, w-1] \end{cases} \tag{2}$$

where the state $k$ represents the current value of the backoff counter of a vehicle. Fig. 2 illustrates that each state $k$, or equivalently backoff value, can be directly selected with probability $1/w$, as shown in the first line of formula (2). Additionally, it is possible to reach the state $k$ by successively decrementing the counter with probability 1 as shown in the second line of formula (2), after the selection of a higher initial backoff value.

From the previous description, it can be concluded that if the backoff counters of two or more vehicles have the same value at the same time, they follow the same decrement way and reach zero at the same time, and then they transmit packets immediately and simultaneously, then the collision will occur. Next, we will model and analyze the probability of this collision.

Assuming there are $n$ vehicles in the system to send data packets. The initial value of backoff counter of each vehicle is an integer, and it is uniformly randomly selected in $[0, w-1]$, as shown in Fig.2. According to the aforementioned counter decrement, to avoid the collision, all the initial values of backoff counters of $n$ vehicles are required to be different from each other, the number of selections of collision-free initial values can be modelled as the combinations without repetition as below

$$C_{w,n} = \binom{w}{n} = \frac{w!}{n!(w-n)!} \tag{3}$$

Formula (3) expresses the number of combinations of $n$ different numbers selected from $w$ numbers.

Whether there is a collision or not, the number of selections of all the possible initial values for backoff counters can be modelled as the combinations with repetition as below

$$H_{w,n} = C_{w+n-1,n} = \binom{w+n-1}{n} = \frac{(w+n-1)!}{n!(w-1)!} \tag{4}$$

Formula (4) denotes the number of combinations of $n$ numbers selected from $w$ numbers, and different from the formula (3), among these $n$ numbers, the cases of same numbers are allowed here.

The collision-free probability of packets sending with contention windows $[0, w-1]$ for $n$ vehicles can be derived from formula (3) and formula (4) below, as shown in formula (5).

$$P_{Collision\ Free} = \frac{C_{w,n}}{C_{w+n-1,n}} = \frac{w!(w-1)!}{(w-n)!(w+n-1)!} \tag{5}$$

The collision probability can be expressed as

$$P_{Collision} = 1 - P_{Collision\ Free} \tag{6}$$

Fig.3 shows collision-free and collision probability, where $w=16$, which is the value of the minimum size of the contention window as per the standard [22]. The number of vehicles in the backoff stage varies from 0 to 15.

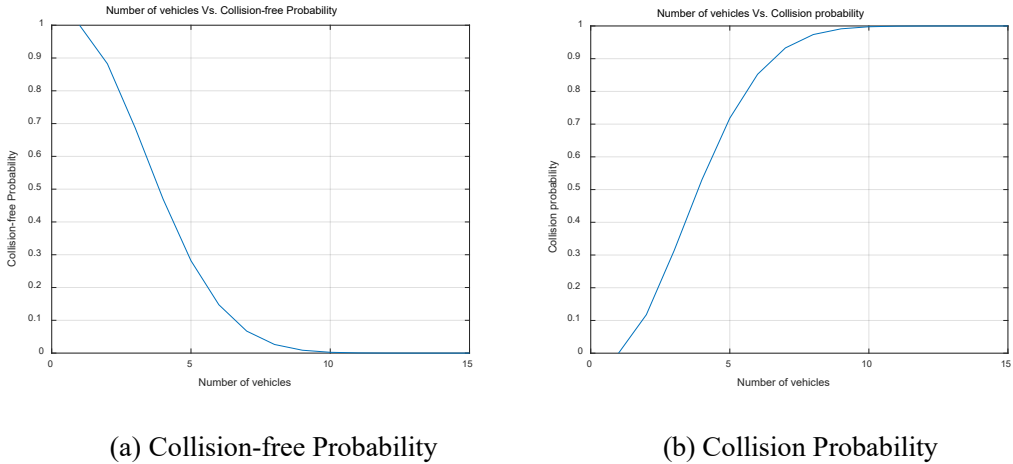

(a) Collision-free Probability      (b) Collision Probability

**Fig.3 Probability with and without collision**

As shown in Fig.3, the collision probability is very sensitive to the number of vehicles in the backoff stage. In Fig.3.(b), the collision probability reaches 71.83% when the number of vehicles increases to 5, however, the collision probability is only 11.76% when there are 2 vehicles in the backoff stage.

### 3.2 Miss My Packets (MMP) Problem

In VANET scenarios, the receiver with only one physical receiving unit can only process one received packet once. It's hence difficult for the receiver to judge beforehand whether the coming packet is desired or not, which might cause the receiver to miss actually desired packets when its receiving unit is occupied by other packets. This situation can be named as miss-my-packets (MMP) problem.

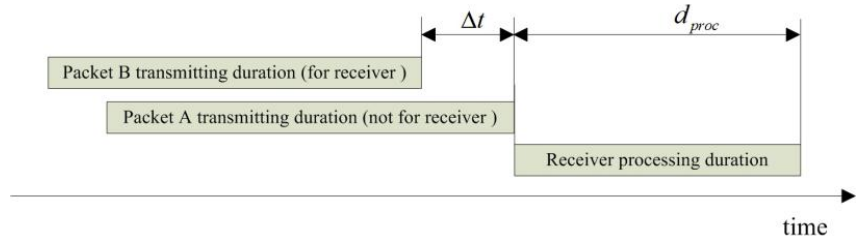

Fig.4 The demonstration of MMP problem

As shown in Fig.4, assuming there are two packets, packet A and packet B, transmitted by two different vehicles through the wireless channel at the same or different frequencies. Packet A reaches the receiver $\Delta t$ earlier than package B, however, the receiver expects to receive packet B rather than packet A. Once packet A is detected, the physical receiving unit of the receiver will start processing packet A and the processing duration is $d_{proc}$. If $d_{proc} > \Delta t$, the receiver will miss the desired packet B and mentioned MMP occurs.

This paper has presented a new approach to alleviate collision issues in CSMA/CA and MMP as mentioned above. From Fig.3 and Fig.4, we can conclude that the fewer vehicles in the backoff phase at the same time, the lower probability of collision and MMP. The proposed approach in this paper is inspired by this investigation.

The novel approach involves the scheduling of transmission time for each link or connection, so it mainly focuses on the lower layer protocol. It is a good point that the lower layer optimization can lead to less overhead. For instance, the time overhead of re-transmission when the MAC layer detects an error is much less than that on the application layer. However, the proposed approach is distinct from that in [17], [18], [20], [27], in which there are a lot of time-slot details involved. Our approach considers scheduling issues on the link level, and avoids complex allocation and synchronisation issues of time slots.

### 4. System Framework and Proposed Algorithm

Similar to the model in [5], the proposed system framework is shown in Fig.5. Vehicles (or nodes) and roadside unit (RSU) are the entities that can send and receive data through wireless channels. The SDN controller is a central controller, which can obtain a global overview of network status by collecting information from fogs and implement network-wide policies. The fog is a subsystem which consists of one RSU-SDN Controller and RSUs within a certain coverage area. The RSU-

SDN Controller can orchestrate RSUs under its coverage, and communicate with RSUs through Openflow protocol [3], [31]. Herein, nodes can directly communicate with others through DSRC links.

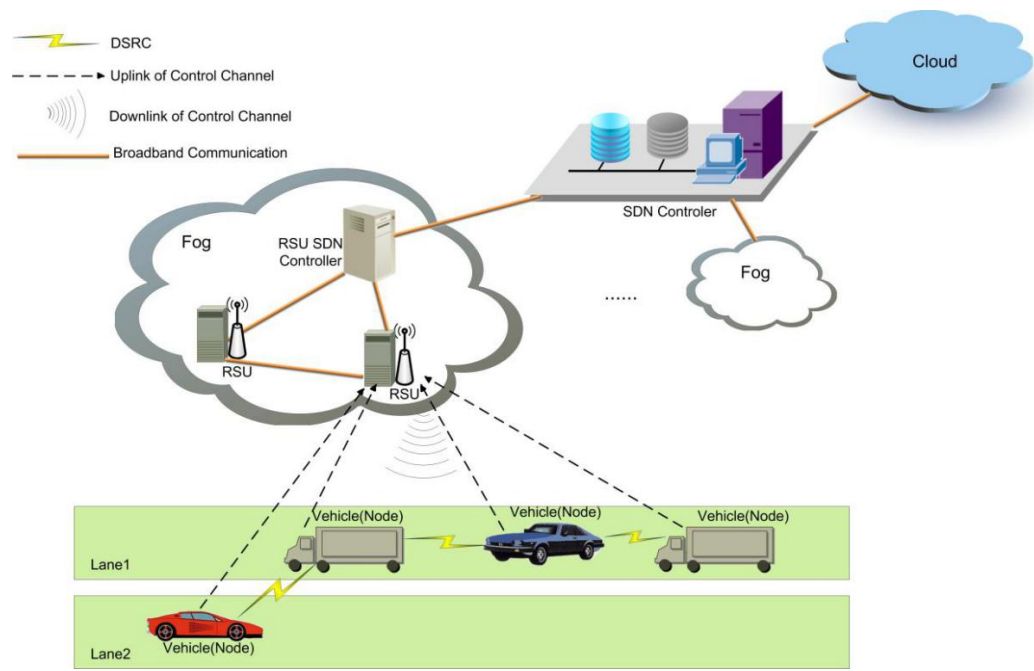

**Fig.5 The proposed system framework**

Assuming that RSU can communicate with nodes through the control channel, and all user data packets are transmitted through DSRC. Obviously, the system works as the following hybrid network:

- Each sender, which intends to send data to other nodes, firstly sends its quality of service (QoS) requirement to RSU through the uplink of the control channel. The QoS requirements include data rate, deadline to complete transmission, delay, etc.

- RSU calculates the total duration of all packets for each connection with consideration of QoS requirements. Then, the controller applies the proposed algorithm to obtain transmission schedule results. Meanwhile, related information is sent back to RSU-SDN controller and further to SDN central controller.

- Through the downlink of the control channel, RSU sends scheduled results back to senders.

- According to the received schedule information, the sender transmits their user data packets to their destination through DSRC links.

As the aforementioned procedure, it is quite important to develop a scheduling algorithm which can both make an efficient schedule and well satisfy the QoS requirement. In this paper, we have proposed an algorithm to fulfil these requirements. In a common SDVN framework, exactly as one in Fig.5, the proposed algorithm can be easily embedded in RSUs as a part of SDVN.

In addition to the end-end delay, the PDR is also an important metric of performance and QoS in VANET. Therefore, our work aims to maximize the PDR of VANET with the least collision and the lowest missing rate, and the objective function can be further expressed as

$$\Omega = \arg\min_{t_i \in T_i} f(\boldsymbol{t}) \tag{7}$$

Subject to $T_i + \tau \leq Q_i, i = 1,2,...,N$

where $T_i$ is the available sending-time window. $Q_i$ is the sending deadline which meets QoS requirement of the $i^{th}$ connection. The connection here means a direct peer-to-peer link between two nodes. $N$ is the number of all connections to be scheduled. $\tau$ is an additional margin for the delay caused by CSMA/CA backoff, without loss of generality, $\tau = 0$ in this paper. $\boldsymbol{t} = \{t_1, t_2, ..., t_N\}$ is a vector consisting of scheduled sending time for each connection, as shown in Fig.6.(a). Each connection involves a sending node and a receiving node, and its duration is the total period during which the sender continuously transmits its data packets over the wireless channel.

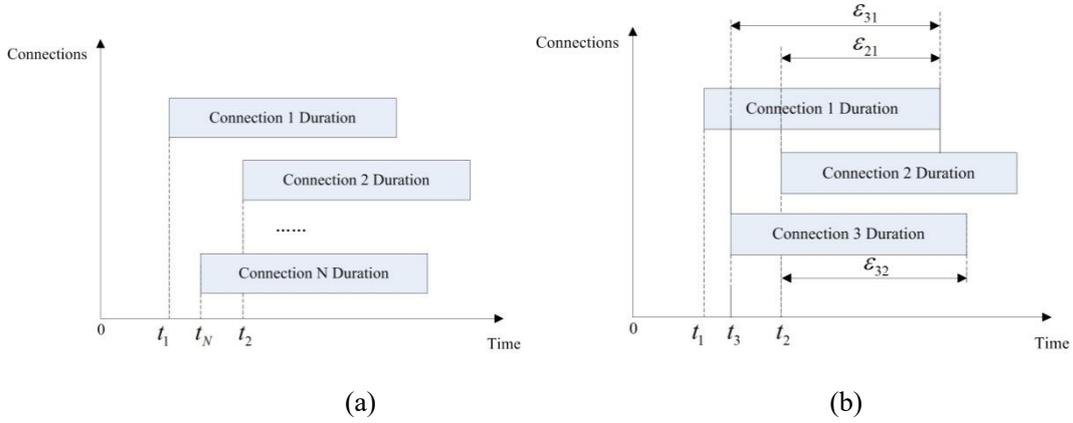

(a)          (b)

**Fig.6 Variables Demonstration**

$f(\boldsymbol{t})$ in formula (7) is a cost function measuring the overlap between durations, and it is defined by

$$f(\boldsymbol{t}) = \sum_{i=1}^{N} \sum_{j=1(j \neq i)}^{N} \varepsilon_{ji} \tag{8}$$

The definition of cost function $f(\boldsymbol{t})$ is based on the fact that less overlap of durations can reduce the collision and uncertain time delay. It is because each duration in Fig.6 consists of multiple packet periods and less overlap of durations means that fewer vehicles simultaneously stay in the backoff stage. As shown in Fig.3, fewer vehicles in backoff stage can reduce collision probability. $\varepsilon_{ji}$ is the time overlap between duration $j$ and duration $i$ as shown in Fig.6.(b).

The time complexity of formula (8) is easily calculated as $o(N^2)$. To get the optimum solution of formula (7), the search time step $\delta$ for $t_i$ must be introduced. According to the size of $\delta$, duration$[0 \quad t_i]$ is divided into $M$ segments, where $M = t_i / \delta$. Given the number of search as $M$, it means each connection has $M$ possible schedules. When $N$ connections are considered in the system, formula (7) has the time complexity $o(M^N)$ which is extremely high. To alleviate the complexity of formula (7), in this paper, we have proposed an algorithm called transmission schedule greedy search (TSGS) as follows.

**Tab.2 Main notations used in the TSGS algorithm**

| Notations | Definitions |
|---|---|
| $N$ | The number of all connections in the system |
| $\boldsymbol{Q}$ | According to the transmission requirements of each connection, define $\boldsymbol{Q}$ as a set of deadline time $\{q_1, q_2, ..., q_N\}$, whose element indicates the deadline time when packets transmissions of each connection should have been completed. |
| $\boldsymbol{W}$ | Set of time windows $\{w_1, w_2, ..., w_N\}$ whose element is the available time period for scheduling. |
| $\boldsymbol{t}$ | Set of scheduled sending time $\{t_1, t_2, ..., t_N\}$ |
| $\boldsymbol{D}$ | Set of connection durations $\{d_1, d_2, ..., d_N\}$ |
| $\varepsilon_{i,j}$ | The size of time overlap between $d_i$ and $d_j$ |
| $\delta$ | Search step of time when scheduling |

In this paper, vector variables are written using bold fonts, and scalar variables are not in bold.

---

**Algorithm:** Transmitting Schedule Greedy Search (TSGS)

---

**Input:** the transmission requirements $\boldsymbol{Q}$

1. **While ( $i \leq N$ )**

    // Get the available time windows $\boldsymbol{W}$ for scheduling, as shown in Fig.7.

2.     $w_i = q_i - d_i$

3. **While ( $i \leq N$ )**

    // $\sigma$ is the number of searching with step $\delta$ for the $i^{th}$ connection

4.     $\sigma = w_i / \delta$

5.     $k = 0$

6.     **While ( $k < \sigma$ )**

    // $J$ is the number of scheduled connections before connection $i^{th}$, so $J = i - 1$

7.         **While ( $j \leq J$ )**

    // Get the cost measurement, i.e. the overlap.

8.         $\Phi_i(k) = \sum \varepsilon_{i,j}(k)$

9.         $k = k + 1$

        // Get the search index with the minimal scheduling cost

10.       $k_{min} = \underset{k \in [0,...,\sigma-1]}{\arg\min}(\Phi_i)$

        // Convert the index to sending time

11.       $t_i = k_{min} \times \delta$

**Output:** $t = [t_1, t_2, ..., t_N]$ // Finished scheduling and obtain sending time for each connection.

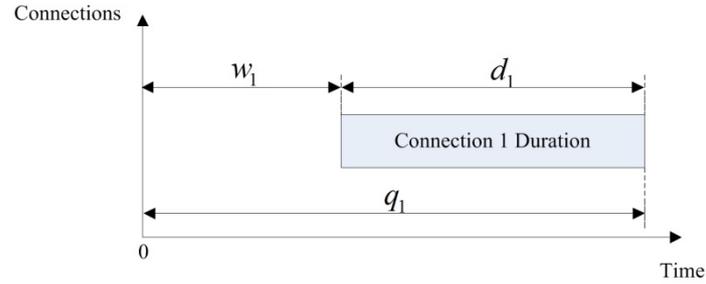

**Fig.7 Time-related variables demonstration**

The proposed TSGS algorithm schedules each upcoming connection $i$ based on the scheduled $J$ connections as the description in the pseudo-code program. Each new connection only considers its $M$ possible schedules and the scheduled connections remain unchanged, therefore the time complexity of the TSGS algorithm is $o(MN)$ which is much lower than $o(M^N)$ obtained by direct searching in formula (7).

## 5. Performance Evaluation

### 5.1 Experimental setup

To investigate the performance of the proposed TSGS algorithm, we conduct extensive simulations on a unified platform including a road network editor for OpenStreetMap, a road traffic generator SUMO, a network simulator Omnet++, and a vehicular network framework Veins. By using OpenStreetMap, we extract the road map of M1 highway near Loughborough University, UK, as shown in Fig.8. The leftmost picture of Fig.8 is the original map captured from OpenStreetMap, whereas the rightmost one is the extracted road map from OpenStreetMap. In this work, we assume all vehicles in the scenario can error-freely communicate with the scheduling controller deployed in RSU.

SUMO is used here to generate a traffic flow, which includes the trajectories of 7 vehicles and generates at most 3 connections requirements. SUMO generates the trajectories of vehicles by following some rules, such as left-hand traffic, speed limitation (45 miles/h), etc. The simulation time is 20 seconds.

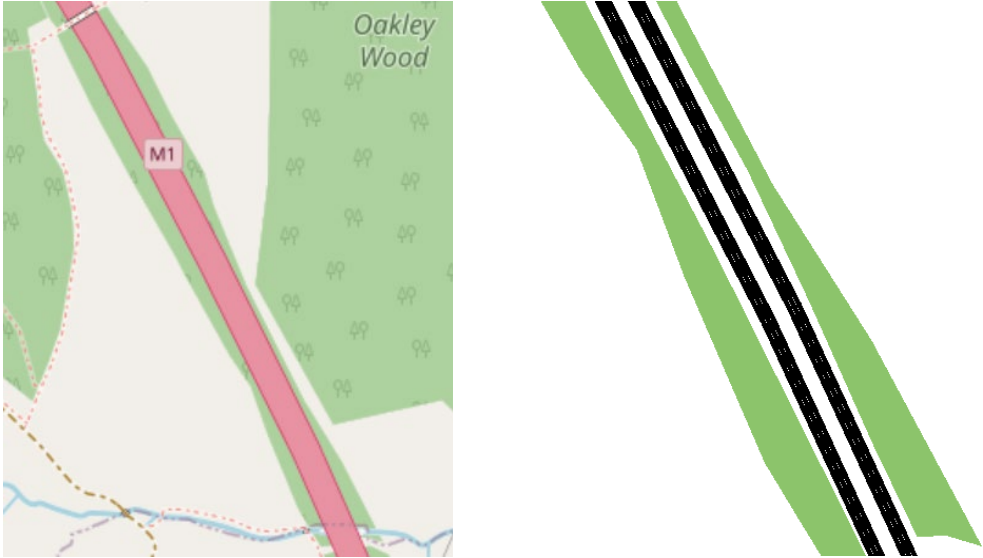

**Fig. 8 M1 highway map near Loughborough University, UK**

The setting of important parameters in OMENT++ for the experiment has been shown in Tab.3.

**Tab.3 The setting of important parameters**

| Parameter | Value |
| --- | --- |
| Transmission Power | 0.5W |
| Speed limit | 45 mph |
| Date rate of DSRC | 6.0 Mbps |
| Bandwidth of DSRC | 10.0 MHz |
| Noise floor | -98dBm |
| Wireless channel model | Simple path loss |
| Frequency of wireless channel | 5.9 GHz |
| DSRC transreceiver for each vehicle | 1 |
| Control signaling transreceiver for each vehicle | 1 |
| Packets for each connection | 50 |
| Time duration of a packet | 23us |
| Scheduling controller (RSU) | 1 |
| Connections | 2 or 3 |
| Vehicles (nodes) | 7 |

## 5.2 Experimental results

### A) The demonstration of the proposed algorithm

In our system, if 3 vehicles are selected as senders and other 3 vehicles as receivers, we can get 3 independent connections each of which consists of a sender and a receiver. The connection here can also be called a link. The number of successive-sending data packets in each connection is set to 50. The QoS requirement keeps the same in all connections, and it is represented as the time window during which the transmissions of all packets must be completed. Herein, the time window is set as [15, 15.0492] second. Fig. 9 demonstrate the basic idea of the proposed TSGS schedule algorithm, where the duration means the total transmission period of 50 packets in each connection.

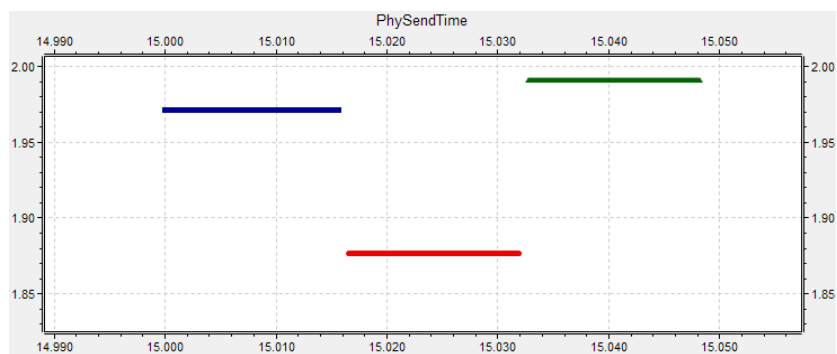

(a) Durations of TSGS schedule algorithm

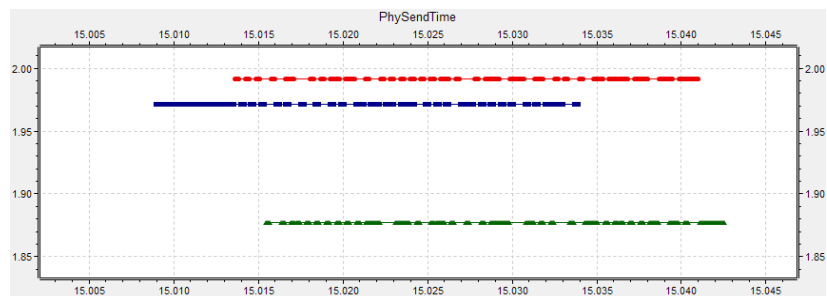

(b) Durations of Random sending

Fig. 9 The duration demonstration of 3 connections

(QoS requirement: Time window to complete transmission is 15s~15.0492s)

Fig.9.(a) demonstrate that the proposed TSGS schedule algorithm can well arrange the transmission for each connection to minimize the overlaps of durations. Alternatively, this mechanism can minimize the probability of activating backoff and reduce the transmission delay for each connection. Meanwhile, the probability of packets collision also decreases.

As shown in Fig.9.(b), in the random sending situation, much duration overlap activates the backoff mechanism of CSMA/CA and causes greater transmission delay for each connection. The backoffs result in the time gaps between packets in each connection being greater than that in Fig.9.(a). In addition, the probability of packets collision increases.

### B) PDR performance

This set of experiments compares the performance of the proposed TSGS schedule algorithm with random sending. Random sending is in line with the actual situation in the distributed system. Since maximizing the PDR or the probability of successfully transmitting packets is the performance objective of this work, we first make the comparison in terms of PDR.

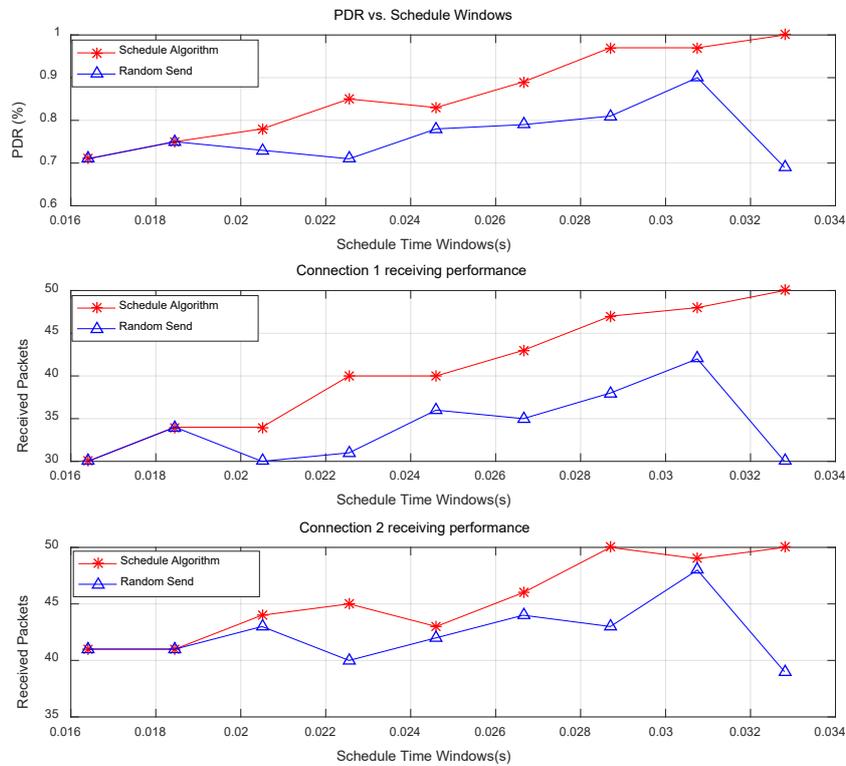

**Fig.10 PDR and received packets with increasing time windows**

**(2 connections)**

Fig.10 shows the total PDR of the system and the number of received packets for each connection when the available schedule time windows gradually increase. Each connection is a random wireless link between two nodes. The available schedule time window of a connection is decided by its QoS requirement, in fact, it is the time window allowed to send packets as $q_1$ shown in Fig.7. Without loss of generality, we assume that the QoS requirements of all connections keep the same in our work, thus each connection has the same schedule time windows. As shown in Fig.10, the proposed TSGS schedule algorithm gives better performance than the random sending mechanism on PDR. As the schedule time window increases, the PDR and the number of received packets in our system increase almost monotonously. However, random sending really cannot have such a performance. In the 2-connections situation, the average PDR of TSGS algorithm is up to 86.11% and 9.78% higher than that of random sending mechanism whose average PDR is only 76.33%. The average PDR is defined as the average value of PDRs under all schedule time windows in the first sub-graph of Fig.10.

That is because as available schedule time window increases, smaller or even no overlap among durations can be achievable by applying the proposed TSGS algorithm as shown in Fig. 9. (a), which reduces the probabilities of both activating backoff and packet collisions.

When the number of connections in the system increases, the probability of packets collision also increases. And the proposed TSGS schedule algorithm will show much better performance. Similarly, in 3-connections situation, , the total PDR performance of proposed schedule algorithm is much superior to that of random sending as shown in Fig.11. In 3-connections situation, the average PDR of TSGS algorithm is up to 95.50% and 15.28% higher than that of random sending whose average PDR is only 80.22%. Meanwhile, it can be easily imagined that our proposed algorithm should be far better than the random sending mechanism in the scenarios with more connections or more complex systems.

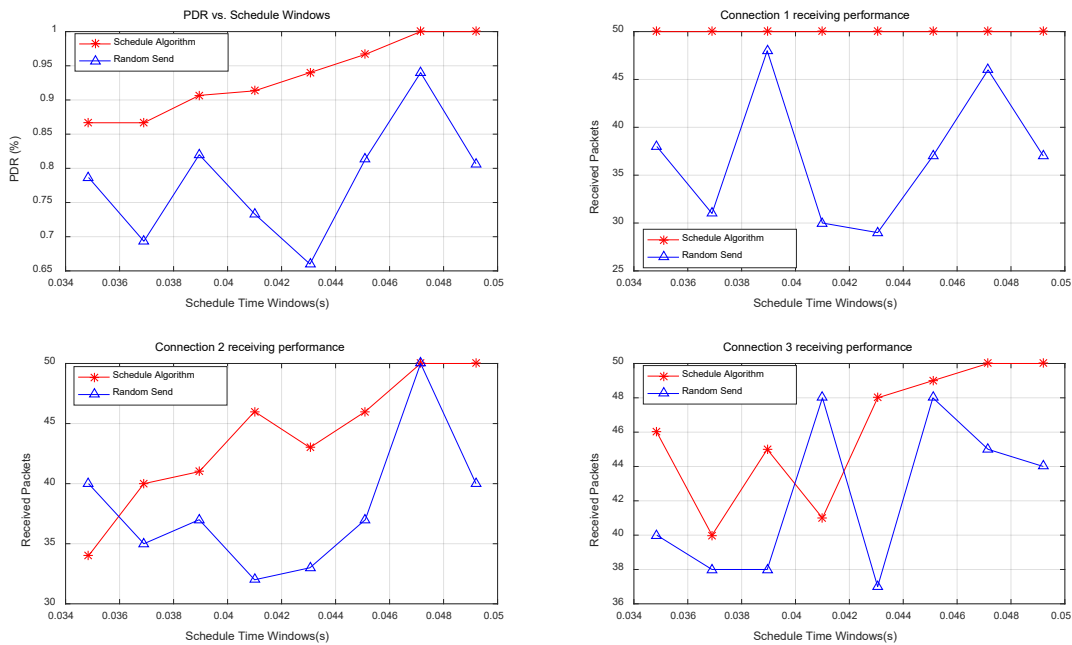

**Fig.11 PDR and received packets comparison (3 connections)**

**C) Packet collision and MMP analysis**

For each vehicle receiving data packets through a connection, it cannot successfully process collided packets in the broadcast situation due to the lack of ACKs. In Fig.12, the collision feature of the proposed algorithm is in comparison with that of random sending. The proposed algorithm can achieve lower collision. Particularly, when the schedule time window reaches 0.0328s, there has been no packet collision for the proposed TSGS schedule algorithm. However, random sending mechanism still struggles with the collision problem. In the 2-connections situation, the average collision ratio of proposed TSGS algorithm is 1.01% and 0.67% lower than that of random sending mechanism whose average collision ratio is 1.68%. Compared with the random sending mechanism, the proposed algorithm improves collision-avoidance performance by about 40%. The average collision ratio is defined as the ratio of the number of all collided packets to the number of all sent packets under all schedule time windows. In addition, it's worth mentioning that one collision involves at least two packets which belong to different connections, therefore, these 2 connections, connection 1 and connection 2, have the same collision performance as shown in last 2 sub-graphs of Fig.12.

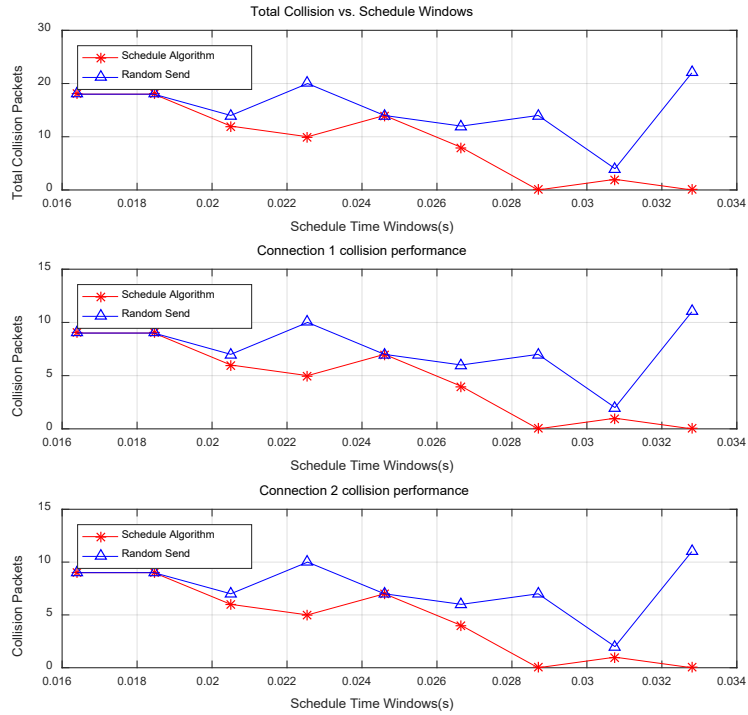

**Fig.12 Collision performance comparison (2 connections)**

Assuming that there is only one DSRC receiving unit for each vehicle in our work, if the receiving unit is processing an undesired packet, however, the physical layer of receiving unit is hard to distinguish it. Meanwhile, another packet actually for the vehicle arrives over the wireless channel, but it cannot be received successfully by the vehicle and will be missed because of the occupied receiving unit. We named this circumstance as miss my packets (MMP). As shown in Fig.13, the proposed schedule algorithm can achieve lower MMP than that of random sending mechanism, which alternatively means higher packet delivery ratio (PDR). For the random sending mechanism, when the schedule time window reaches 0.0328s, the MMP numbers of both connection 1 and connection 2 jump to 20, meanwhile there is no MMP for the proposed algorithm.

In the 2-connections situation, the average MMP ratio of random sending is 6.57% and 2.97% higher than that of the proposed algorithm whose average MMP ratio is close to 3.6%. The average MMP ratio is defined as the ratio of the number of all MMP packets to the number of all sent packets under all schedule time windows in Fig. 13.

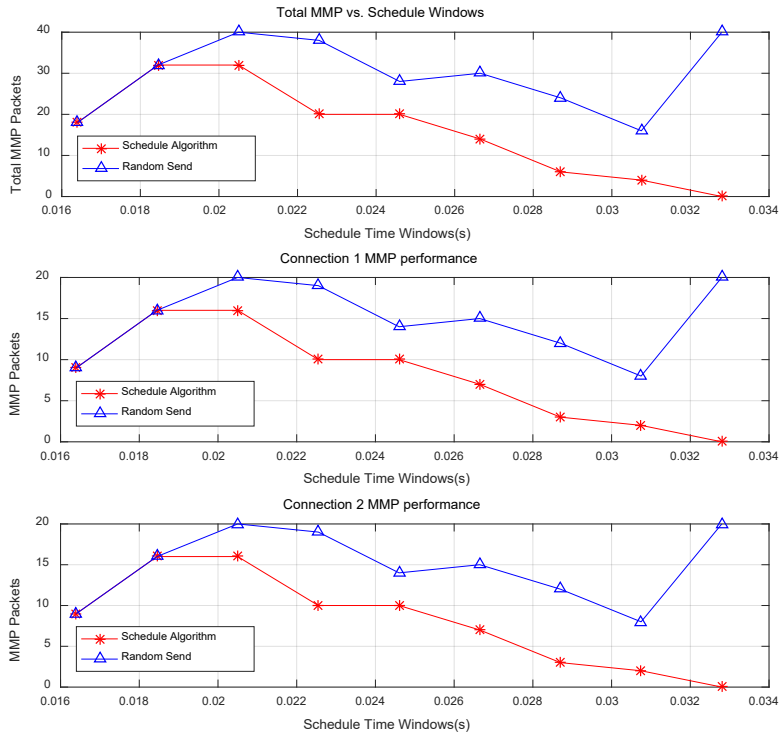

**Fig.13 Miss my packets performance (2 connections)**

Next, we would analyze the relationship among PDR, collision and MMP. For example, when the schedule time window reaches 0.0328s, in the random sending situation, connection 2 obtain 39 successfully delivered packets as shown in Fig.10, which alternatively means this connection has missed 11 packets. It's also consistent with the 11 collided packets shown in Fig.12. However, the number of MMP packets for connection 2 is 20 under the 0.0328s time window as shown in Fig.13, which is larger than the number of missed packets. The reason is that when receiving unit is occupied, new arriving packets will be counted as MMP packets no matter whether it is for the vehicle or not.

### D) QoS and delay analysis

Here we will analyze the delay characteristics of the proposed scheduling algorithm. As shown in Fig.14, 3 connections are considered in the analysis. For each sub-graph, the dotted line indicates the QoS-related time requirement to complete the transmission. In fact, in our simulation setting the QoS-related time requirement is the same as the available schedule time window, so it can be seen that the QoS time requirement is a straight line that slopes upward as the schedule time window increases. For connection 1, the time to complete transmission by applying the proposed schedule algorithm is always less than the QoS requirement, which means the proposed schedule algorithm perfectly meet the QoS requirement. Meanwhile, the time is always less than that of the random sending mechanism. For connection 2, when the schedule time windows are greater than 0.038 s, the transmission time of the proposed algorithm is less than the required QoS-related time. For connection 3, the threshold is about 0.046s, when the available schedule windows exceed the threshold, the time to complete transmission by applying the proposed algorithm also completely meet the QoS requirement. However, the random sending mechanism has not this threshold characteristic.

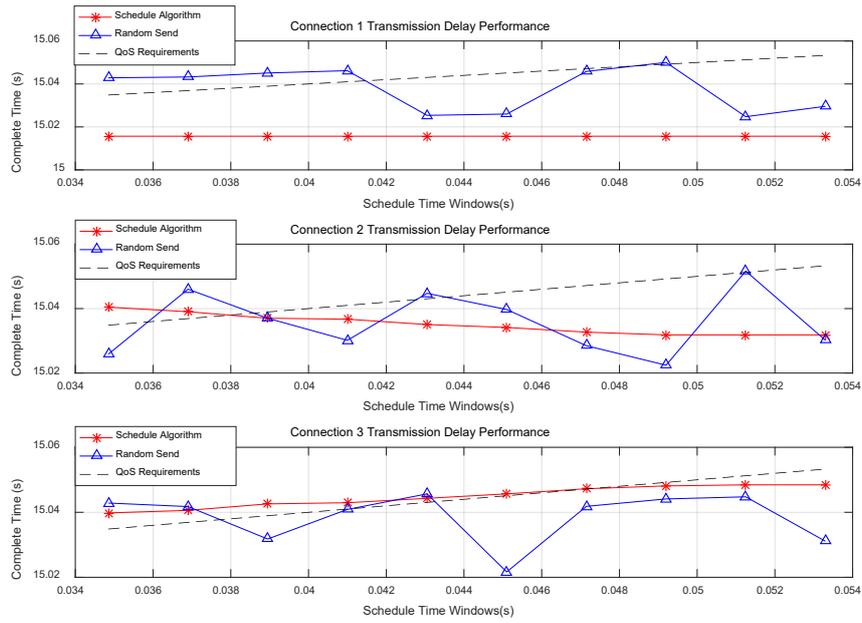

**Fig.14 Delay performance comparison**

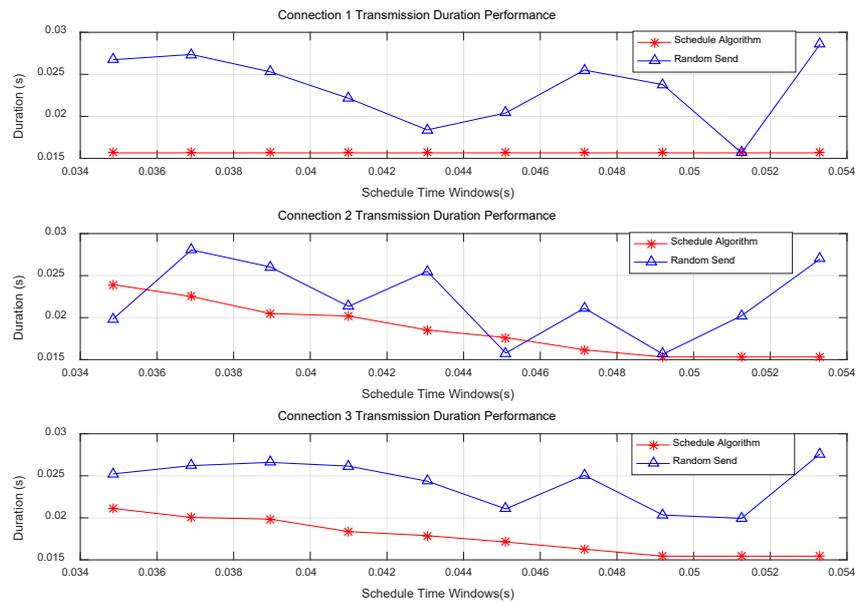

**Fig.15 Transmission duration comparison**

Fig.15 shows time spent by two different scheduling methods to transmit the same number of data packets under different time scheduling windows. The calculated time is the time period from the start-sending of the first data packet to the transmission completion of the last data packet. As shown in the figure, in almost all cases, the proposed schedule algorithm takes less time than the random sending mechanism. By comparison, the proposed algorithm reduces the transmission durations by 33.04%, 15.92% and 27.04% for the three connections respectively. Of course, the absolute time delay to complete the transmissions needs to consider the start time, as shown in Fig.14. However, Fig.15 indicates that the proposed schedule algorithm leads to fewer backoffs than the random sending mechanism.

## 6. Conclusions

In this paper, we introduce the problems probably encountered by the distributed VANETs. Due to the lack of global information and limited receiving units, the packets collision and miss-my-packets (MMP) problems will arise. We have theoretically analyzed the collision probability of CSMA/CA in the broadcast situation and analyzed the reason for MMP problem in the limited receiving unit. Based on the analysis, we conclude that fewer vehicles in the backoff phase at the same time can lead to a lower probability of collision and less MMP. It is further concluded that reducing the overlap between connections can alleviate collision and MMP when each connection consists of multiple consecutive packets transmissions. Then under the SDVN structure, an innovative link-level scheduling strategy is proposed to schedule the transmission of all connections. A related algorithm named TSGS is also developed to find the optimal sending time for each connection with higher PDR, meanwhile, meet the QoS requirement in terms of transmission delay. The VANET scenarios with 2 connections and 3 connections are built on our unified platform including OpenStreetMap, SUMO, Omnet++, and Veins. Based on this platform, extensive simulation experiments have been conducted to compare the proposed algorithm with the random sending mechanism. The experimental results demonstrate the superiority of the proposed algorithm with 15.28% higher PDR than the random sending mechanism in the 3-connections situation. Meanwhile, the proposed algorithm improves collision-avoidance performance by about 40% and reduce the MMP by 3.6% in the 2-connections situation. Moreover, the proposed algorithm is easily embedded in common SDVN architecture and seamlessly integrated with VANET routing schemes.

## Acknowledgement

Authors thank the funding of Science and Technology Program of Sichuan Province (2018JY0507), China.